\documentclass[doublecol]{epl2} 

\usepackage{amsmath}
\usepackage{amssymb}
\usepackage{amsfonts}
\usepackage{ulem}

\DeclareSymbolFont{matha}{OML}{txmi}{m}{it}
\DeclareMathSymbol{\varv}{\mathord}{matha}{118}

\title{Rugotaxis: Droplet motion without external energy supply}
\shorttitle{Rugotaxis} 

\author{Panagiotis E. Theodorakis\inst{1} \and Sergei A. Egorov\inst{2,3,4} \and Andrey Milchev\inst{5}}
\shortauthor{Theodorakis \etal}

\institute{                    
  \inst{1} Institute of Physics, Polish Academy of Sciences, Al. Lotnik\'ow 32/46, 02-668 Warsaw, Poland \\
  \inst{2} Department of Chemistry, University of Virginia, Charlottesville, VA 22901, USA \\
  \inst{3} Institut f\"ur Physik, Johannes Gutenberg Universit\"at Mainz, 55099 Mainz, Germany \\
  \inst{4} Leibniz-Institut f\"ur Polymerforschung, Institut Theorie der Polymere, Hohe Str. 6, 01069 Dresden, Germany \\
  \inst{5} Bulgarian Academy of Sciences, Institute of Physical Chemistry, 1113 Sofia, 
Bulgaria \\
}

\abstract{
Nano-patterned substrates offer possibilities for controlling the motion of fluids without external
energy supply in novel technologies in microfluidics, coatings, \textit{etc.} Here, we report
on the rugotaxial motion of droplets on wrinkled substrates with gradient in the wavelength
of the wrinkles by exploring a broad range of parameters, such as amplitude of the wrinkles,
substrate wettability, droplet size and wavelength gradient. Adopting a theoretical and
molecular dynamics approach, we determine the Cassie--Baxter and Wenzel states of the
droplets, investigate the efficiency of rugotaxis as a function of different parameters,
and discuss additional effects, such as pinning.
We find that shallow wrinkles characterised by 
small wavelength gradients, and moderate adhesion of the droplet to the substrate
favour the rugotaxis motion with growing droplet size, when pinning is avoided.
We also find that the driving force in rugotaxis is the gain in
interfacial energy between the droplet and the substrate as the droplet enters regions of
denser wrinkles (smaller wavelengths of the wrinkles).}

\begin{document}

\maketitle

The development of various technologies in microfluidics, microfabrication, coatings, 
and biology require the motion of fluids along predetermined trajectories \cite{Srinivasarao2001,Chaudhury1992,Wong2011,Lagubeau2011,Prakash2008,Darhuber2005, Yao2012, Li2018}.
A possibility of realising such motion is by using \textit{gradient} substrates,
namely substrates with gradually changing properties in a certain direction along the
substrate \cite{Bardall2020,Karapetsas2016,Zhang2019,Leng2020}. 
For example, by exploiting differences in tissue stiffness, cells are able to move from
softer to stiffer regions, a phenomenon known as durotaxis \cite{Lazopoulos2008,Palaia2021}. 
Durotaxis is particularly appealing for nanotechnology applications, because fluid motion
is sustainable without providing external energy from a source \cite{Barnard2015,Chang2015,Theodorakis2017,Becton2016}.
In contrast, motion caused, for example, by a temperature gradient (thermotaxis) would
require external energy supply into the system to maintain the gradient that 
is responsible for the fluid motion \cite{Becton2014}. 
In this regard, characteristic cases of droplet motion caused by external
energy supply are chemically driven droplets \cite{Santos1995,Lee2002}, and droplets
on vibrated substrates \cite{Daniel2002,Brunet2007,Brunet2009} or wettability
ratchets \cite{Buguin2002,Thiele2010,Noblin2009}.

Hiltl and B\"oker have recently demonstrated in their experiments the possibility of
causing the motion of a water droplet onto a sinusoidal wrinkle-patterned, solid substrate
without using an external energy source, a phenomenon known as rugotaxis \cite{Hiltl2016}. In this
case, the wavelength characterising the wavy shape of the wrinkles decays as a function of
the position, namely, there is a wavelength gradient characterising the wrinkles \cite{Hiltl2016}. 
Hiltl and B\"oker have found that the droplet moves toward smaller wrinkle dimensions (wrinkles
described by smaller wavelength). By investigating wrinkled substrates with wavelengths
between 230 and 1200 nm, and amplitudes ranging from 7 to 230 nm, they attributed this
phenomenon to the imbalance of the receding and advancing contact angles. 
In particular,
contact angles correlate with the wavelength of the wrinkles, that is, larger contact angles
for larger wavelengths, which naturally corresponds to differences of the substrate density
along the direction of the wavelength gradient and the rugotaxial motion. 
The imbalance between the advancing and the receding contact angles 
as the driving force for droplet motion
has been theoretically discussed in detail by Brochard \cite{Brochard1989} in the 
context of chemical or thermal gradients.
Recently, an asymptotic theory has been developed to match the
advancing and the receding angles to respective solutions of the problem
at the microscale, using the velocity as a small parameter of an
asymptotic expansion, which determines droplet shape and velocity as a
function of the wettability gradient \cite{Pismen2006}.

While wrinkled substrates without gradients \cite{Cerda2003, Republic2020,Hiltl2011,Lee2013,Sun2016,Chung2011,Jeong2010,Carbone2005} and substrates of similar geometries \cite{Ko2015,Ambrosia2015,Spori2008,Kumar2013,Grzelak2010,Liu2020,Liu2021} 
have been studied in different contexts, rugotaxis have remained unexplored. Inspired by the
work of Hiltl and B\"oker\cite{Hiltl2016}, we employ theoretical \cite{LoVerso2011,Milchev2018} and molecular dynamics (MD) \cite{Theodorakis2017}
modelling to investigate the self-propelled motion of nanodroplets on wrinkled, solid
substrates with wavelength gradient characterising the wrinkles. We show that the efficiency
 depends on the particular choice
of parameters for the substrate, which leads to specific scenarios, when rugotaxis is
possible. We also find that the driving force of the rugotaxis phenomenon is the gain in the
interfacial energy between the droplet and the substrate as the droplet reaches areas of
denser wrinkles. Toward studying rugotaxis, we have also explored the transition between
Cassie--Baxter \cite{Cassie1944} (CB, a state where the liquid droplet does not fully penetrate the
grooves on rough surfaces and leaves air gaps between the droplet and
the substrate) and  Wenzel \cite{Wenzel1936} (W, a state where the droplet fully
penetrates the grooves) states in
substrates without gradient and compared our results with theoretical predictions
\cite{Carbone2005} in order to set the stage for a better understanding of 
rugotaxis.

\begin{figure}[t!]
 \includegraphics[width=0.49\textwidth]{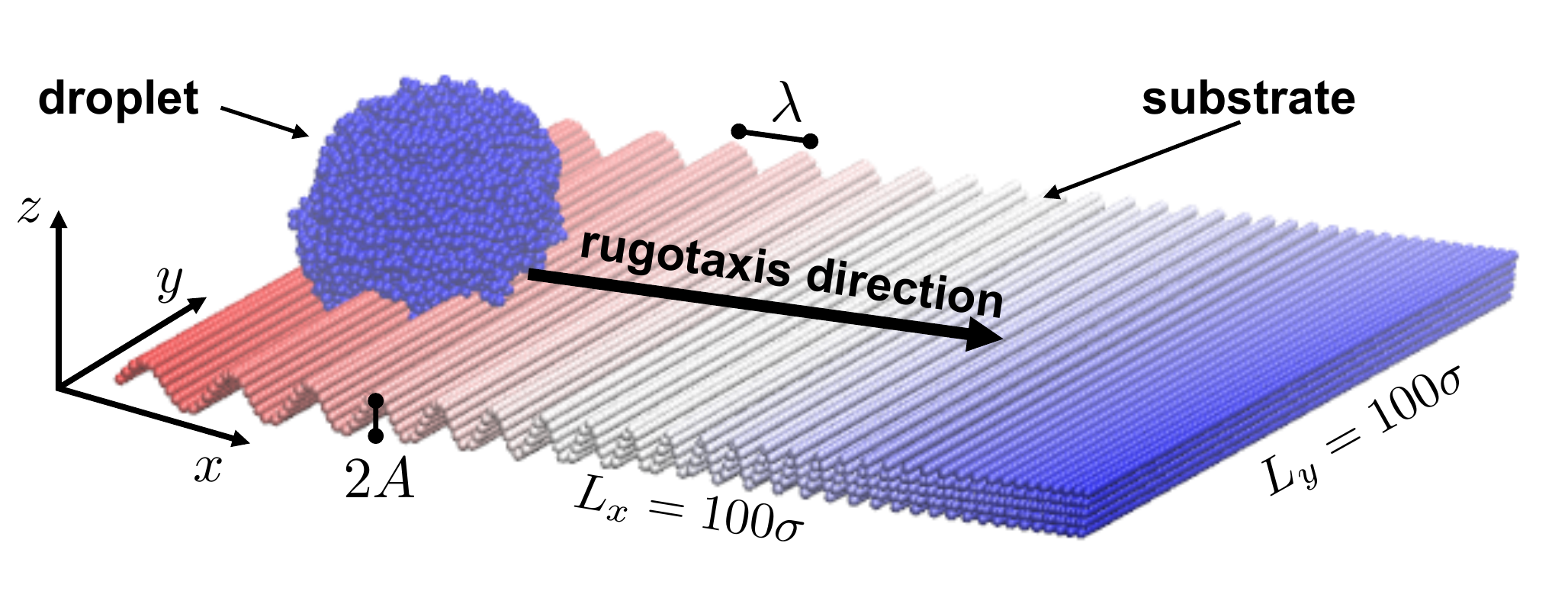}
\caption{\label{fig:1} A typical snapshot of an initial configuration for the rugotaxis
study with a droplet on a wrinkled substrate with gradient in the wavelength, $\lambda$,
characterising the sinusoidal shape of the wrinkles. The dimensions of the substrate in the
$x$ and $y$ directions are $L_x = L_y = 100 \sigma$. The amplitude of the wrinkles is
$A=2\sigma$,  the initial wavelength at the very left side of the substrate is
$\lambda_0=10\sigma$ and that at the right side of the substrate is $\lambda_E=\sigma$.
Given the linear decrease of the wavelength in the $x$ direction, the wavelength gradient 
is constant and equal to $G_\lambda=(\lambda_0-\lambda_E)/L_x = 0.09$, in this case. The
rugotaxial motion of the droplet takes place in the $x$ direction, along the wavelength
gradient, as indicated by an arrow. Snapshots have been produced using VMD software
\cite{Humphrey1996}.}
\end{figure}

Figure~\ref{fig:1} illustrates an initial configuration of our \textit{in silico}
experiments for a particular case. A liquid droplet is placed on a sinusoidal substrate with
wavelength gradient of the wrinkles in the $x$ direction. This requires that the $x$ and $z$
coordinates of the beads be associated with the relation $z = A\sin{(2\pi x / \lambda)}$, 
where $A$ is the amplitude and $\lambda$ is the wavelength characterising the wrinkles.
To implement a gradient in the wavelength of the 
wrinkles, one needs to multiply the wavelength with a coefficient that depends 
linearly on the position $x$ of the bead on the substrate.
Such substrate design is able to
cause the rugotaxial motion of the droplet in the $x$ direction along the wavelength
gradient. Rugotaxial motion will take place for a particular set of parameters
characterising the substrate design, such as the amplitude of the wrinkles, $A$, as well as
the initial wavelength, $\lambda_0$ at the one end of the substrate and the final wavelength
at the other end, $\lambda_E$ ($\lambda_0 > \lambda_E$). The latter wavelengths
determine the wavelength gradient of
the wrinkles, expressed as $G_\lambda=(\lambda_0-\lambda_E)/L_x$ ($L_x$ is the linear
dimension of the substrate in the $x$ direction). The adhesion of the
droplet to the substrate is controlled through the $\varepsilon_{sp}$ parameter of the
Lennard-Jones potential. The details of our MD model are the same as in Ref. \cite{Theodorakis2017} (see, also, Supplementary Information). In order to suppress droplet evaporation, we consider polymer melt droplets made of sufficiently long linear 
macromolecules of length $N=10$ monomers each.
This choice ensures the absence of evaporation effects, which might
have otherwise affected the outcome of our \textit{in silico} experiments \cite{Theodorakis2017}.
In our study, different droplet sizes were considered, namely $N_p=100$, $600$,
and $4800$, where $N_p$ is the total number of polymer chains comprising the droplet. 

\begin{figure*}
\centering
  \includegraphics[width=0.9\textwidth]{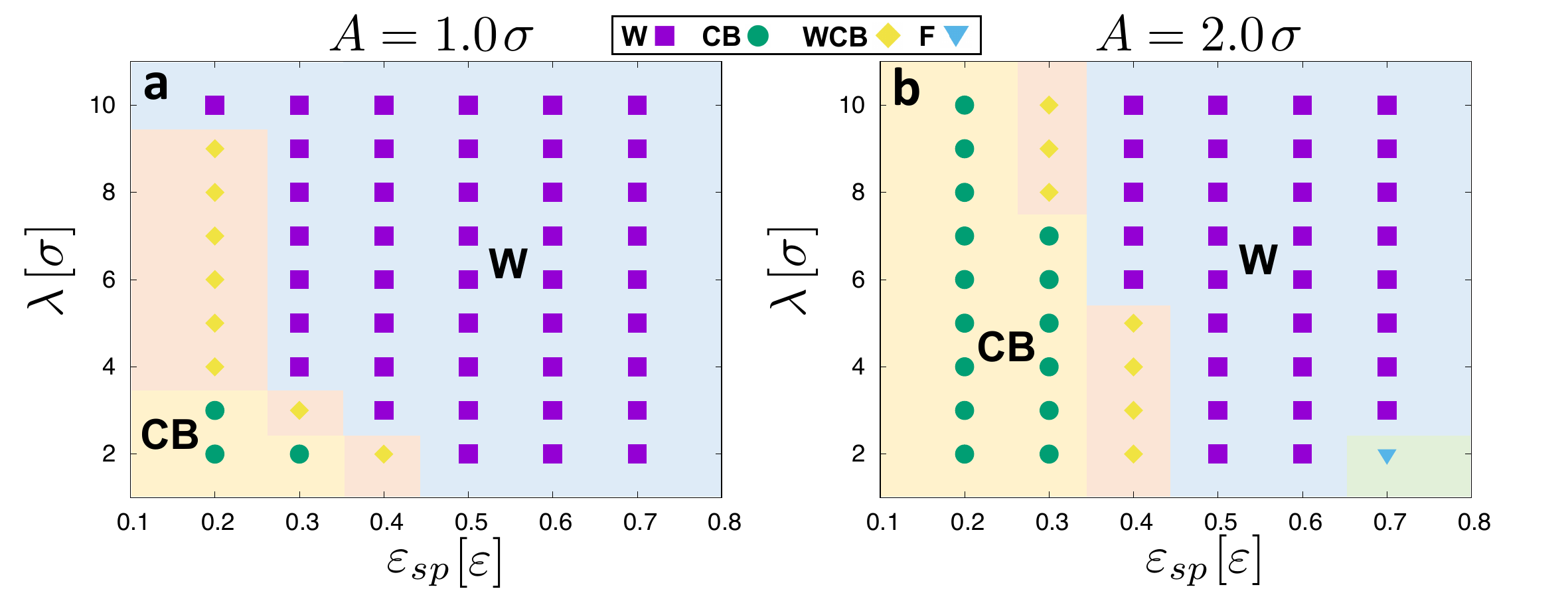}
\caption{\label{fig:2} State diagrams for droplets on wrinkled substrates without gradient
as a function of the wavelength, $\lambda$, and the substrate--droplet attraction strength,
$\varepsilon_{sp}$, for different substrate amplitudes, as indicated. Symbols are as
follows:  W: Wenzel state, CB: Cassie–-Baxter state, WCB: droplet is in the W or the CB
states at different instances in time during the simulation due to the 
thermal fluctuations, and F: Film (droplet spreading due
to the strong attraction between the droplet and the substrate as the density of the
wrinkles increases). Here, $N_p=600$.
}
\end{figure*}

By investigating a range of amplitudes ($A=0.5$--2.5$\sigma$), wavelengths
($\lambda=1$--10$\sigma$) and strengths of attraction
($\varepsilon_{sp}=0.1$--0.7$\varepsilon$) between the droplet and the substrate,
we have initially determined the state [\textit{e.g.} Wenzel (W), Cassie--Baxter (CB)] of
the droplet on substrates \textit{without gradient} by using MD simulation. Our results are presented
in the form of state diagrams in Fig.~\ref{fig:2} for two cases, namely $A=\sigma$ and
$A=2\sigma$. The cases $\lambda=\sigma$ are not shown in the figure, because they represent flat
substrates and a discussion about W and CB states would be irrelevant. Also, cases that the
attraction strength $\varepsilon_{sp}=0.1\varepsilon$ were omitted, because the droplet would
detach from the substrate due to its thermal fluctuations. The results of Fig.~\ref{fig:2} are in
agreement with theoretical predictions \cite{Carbone2005}, which suggest that the CB state
gradually appears in our diagrams in place of the W state as the amplitude of the wrinkles
increases, which translates into the increase of the substrate roughness $A/\lambda$ for each
$\lambda$. Moreover, an increasing strength of attraction between the droplet and the
substrate favours the W state, in agreement with the theory \cite{Carbone2005}. While for
very small amplitude ($A=0.5\sigma$) the W state appears even for very small strength of
attraction between the substrate and the droplet and the CB state is only partially present
during the simulations, for $A>0.5\sigma$, the CB state becomes stable and appears in a larger
number of cases as the roughness $A/\lambda$ increases. However, the MD results also
indicate the existence of a metastable state (WCB; Figure~\ref{fig:2}), where the droplet is
either in the W or the CB state during the simulation. In addition, we observe that the WCB
regime is generally narrow and, also, absent in particular cases as $\varepsilon_{sp}$
increases.  Our results can  be summarised as follows: The formation of the W state is
favoured by a small  amplitude, $A$, a large wavelength, $\lambda$, and a large attraction
between the  droplet and the substrate, $\varepsilon_{sp}$. Our results are in agreement
with the theoretical predictions and further comparisons with theory will be discussed in
the following \cite{Carbone2005}.

\begin{figure*}
\centering
  \includegraphics[width=\textwidth]{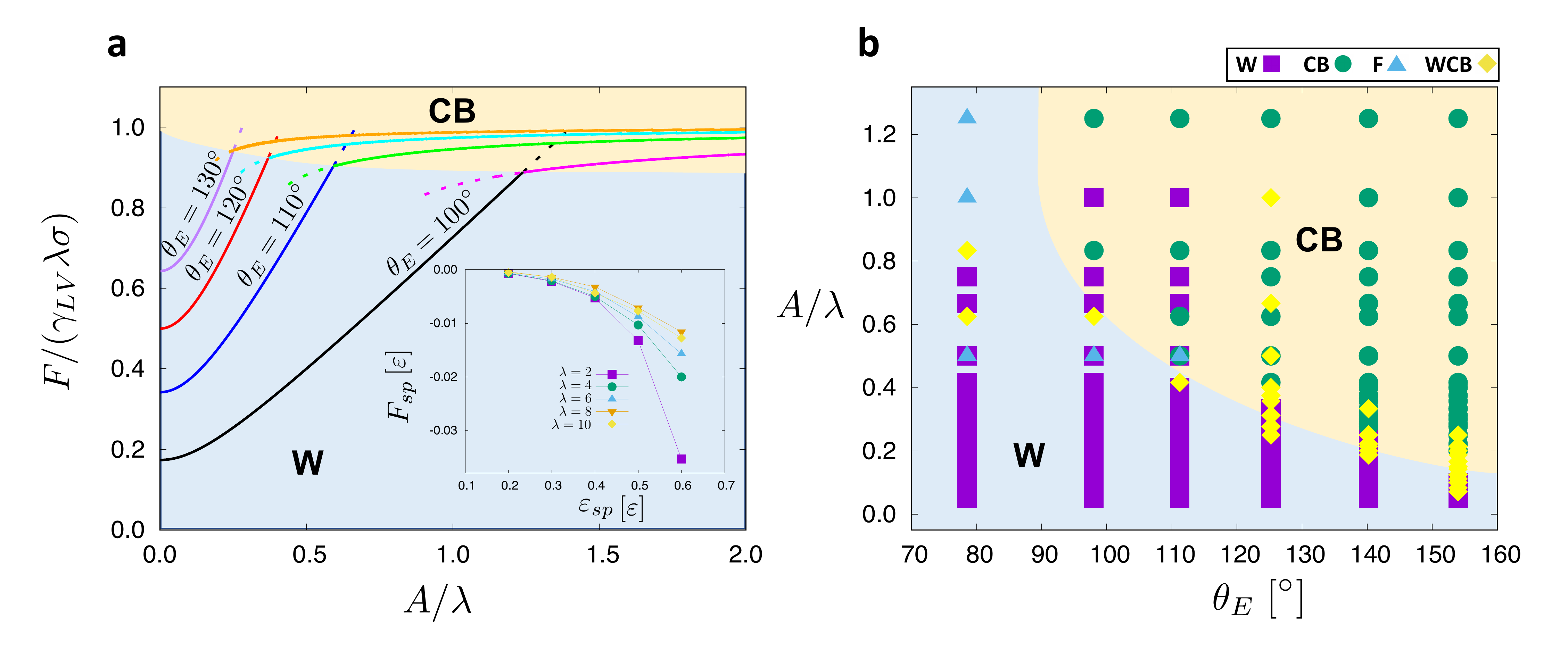}
\caption{\label{fig:3} (a) Free energies of Wenzel (W) and Cassie--Baxter (CB) states of a
droplet with $N_p=600$ as a function of substrate roughness, $A/\lambda$, for four values of
the contact angle $\theta_E$, as indicated. Solid lines correspond to stable states while
dashed lines correspond to the metastable states. Inset presents the free energy component
of the interaction between the droplet and the substrate as a function of the interaction
parameter $\varepsilon_{sp}$ for a substrate with $A=2.0\sigma$. (b) Diagram of different states
(W, CB, WCB, F) as a function of the roughness, $A/\lambda$, and the contact angle,
$\theta_E$ (cf. Fig.~S1 of Supplementary Information showing the
theoretical prediction\cite{Carbone2005}).
}
\end{figure*}

The equilibrium contact angle \cite{deGennesbook,DeGennes1985,Theodorakis2015,Smith2018},
$\theta_E$, provides the means of theoretically computing the
free energies of Wenzel (W) and Cassie--Baxter (CB) states as a function of the substrate
roughness $r=A/\lambda$ \cite{Carbone2005}. These results are presented in
Fig.~\ref{fig:3}a for four representative values of $\theta_E$. The free energies are
given in dimensionless form, $F/(\gamma_{LV}\lambda\sigma)$, where $\gamma_{LV}$ is the
surface tension of the liquid--vapour boundary, and $\sigma$ is the monomer
diameter. As expected (Fig.~\ref{fig:2}a), with increasing roughness, the W state becomes
metastable, while the CB state becomes stable. For each value of $\theta_E$, this transition
occurs at a particular critical roughness, $A_c/\lambda$. Above the line, the CB states are
stable, while below the line the W states are stable. Following this analysis, the results
obtained from the simulation are presented in Fig.~\ref{fig:3}b, which are in very good
agreement with the theoretical predictions (see Fig.~S1 and discussion in Supplementary Information). In particular, for $\theta_E<90^\circ$, only W
states are possible. In addition, as $\theta_E$ approaches 180$^\circ$, even a small degree of
roughness can induce the transition from W to CB states. The simulation results indicate
that the boundary between the W and CB becomes sharper for larger values of $\theta_E$.
Finally, the CB state is absent for $\theta_E<90^\circ$ ($\varepsilon_{sp}>0.6\varepsilon$), in
agreement with the theoretical calculations \cite{Carbone2005} (Fig.~\ref{fig:3}a). We have
also analysed the different contributions to the free energy on the basis of a hybrid MD--DFT
approach \cite{LoVerso2011,Milchev2018} (see Supplementary Information for 
more details on the method and references therein). 
In particular, we have identified the attractive term between the droplet and the
substrate, $F_{sp}$, as the main factor that determines the state of the droplet (W or CB).
Interestingly, the dependence of $F_{sp}$ on the attraction strength, $\varepsilon_{sp}$,
shows a faster than linear decrease as $\varepsilon_{sp}$ increases, that is when the adhesion of the droplet to the
substrate is stronger. This is due to the
increase of the substrate density as the wrinkles become denser for smaller wavelengths,
$\lambda$ (inset of Fig.~\ref{fig:3}a). Moreover, differences in  $F_{sp}$ are small, at
least for weaker adhesion, when the droplet is in `loose' contact with
the substrate.

We have analysed the rugotaxial motion for a wide and
relevant range of parameters. From our results, we have determined that each case is
distinct. For example, two substrates with the same gradient, $G_\lambda$, but different
sets of ($\lambda_0$, $\lambda_E$) can show completely different behaviour regarding their
ability to cause rugotaxial motion. Apart from these obvious differences, certain
parameter combinations can cause the droplet to be in the W or CB states, the formation
of a film, pinning, and others. 
In a recent molecular dynamics study \cite{Theodorakis2021}, 
it has been found that even the slightest heterogeneity on the substrate 
can hinder the motion of a droplet on a substrate (pinning). 
Overcoming such a pinning barrier (\textit{i.e.} depinning) can take place by
increasing the interfacial energy between droplet and substrate due to
changes of the substrate properties along the direction of motion. Thus,
the difference in wettability on both sides of the pinning barrier has
been exploited \cite{Theodorakis2021}, while in our study the droplet steps over the
barrier due to the decrease of the wavelength, which in turn results in
an increase of the interfacial energy between droplet and substrate.
This further highlights that successful and efficient rugotaxis
requires an appropriate design, which takes into account all these effects. 

\begin{figure*}
\centering
  \includegraphics[width=0.9\textwidth]{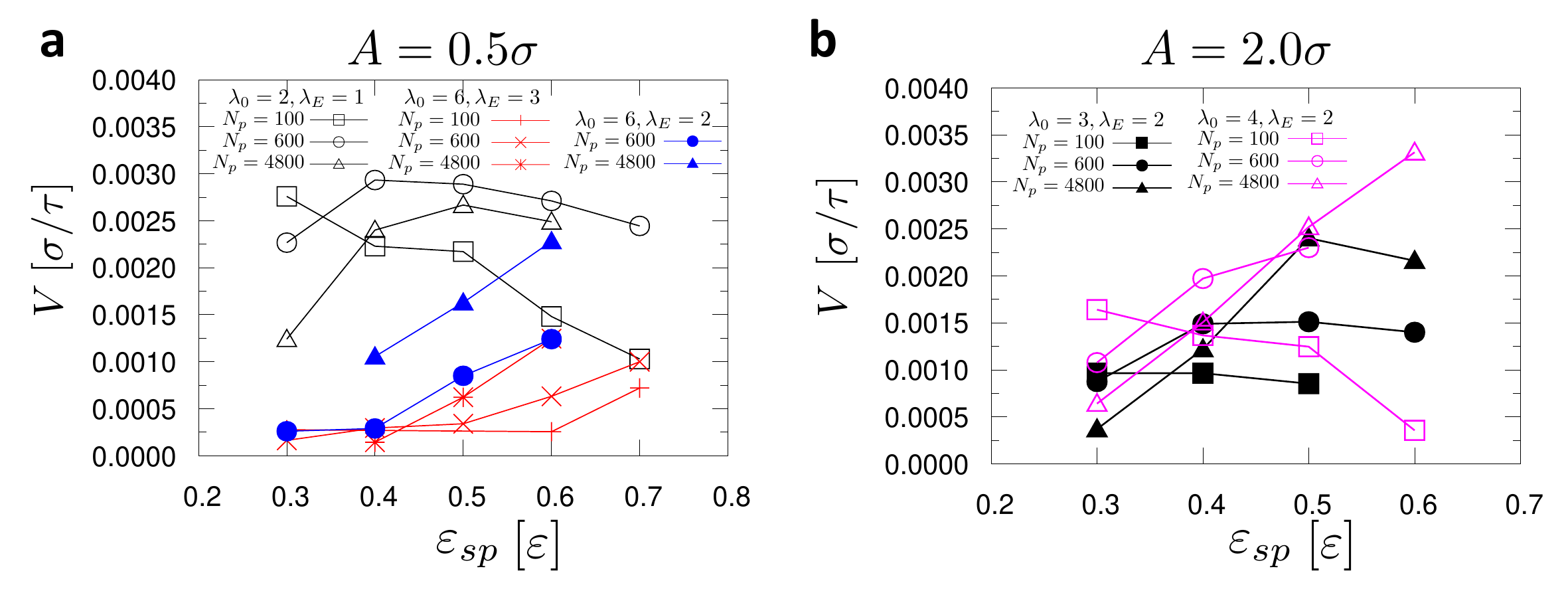}
\caption{\label{fig:4} The dependence of the average rugotaxis velocity, $V$, on the
attraction strength, $\varepsilon_{sp}$, for different cases of $\lambda_0$,
$\lambda_E$, $N_p$ and $A$, as indicated.
}
\end{figure*}

Based on our previous experience with investigations on the durotaxis phenomenon
\cite{Theodorakis2017} and the experience gained by the current study, we have determined
that the average rugotaxial velocity, $V$, of the droplet is an appropriate measure to
assess the overall efficiency of the rugotaxis motion in successful cases, that is cases
that can uninterruptedly translocate throughout the substrate in the direction of the 
wavelength gradient. In this case, $V=L/t$, 
where $L$ is the distance covered by the centre of mass of the droplet during its 
translocation from the very left part of the substrate to
the very right part of the substrate in the $x$ direction (Fig.~\ref{fig:1}), 
and $t$ is the time required to complete the rugotaxis motion. Then, Fig.~\ref{fig:4}
presents various characteristic examples that illustrate
the dependence of the velocity for different droplet size and different
scenarios of successful rugotaxis cases.
In general, larger droplets are able
to easier overcome the pinning barriers \cite{Theodorakis2021}, with the latter becoming more
pronounced as the droplet is in the Wenzel state (Fig.~\ref{fig:2}), which otherwise favour
the rugotaxial motion. Still, one can observe the difference
in the behaviour of large droplets ($N_p=4800$) between substrates with $A=0.5\sigma$ and 
$A=2.0\sigma$, where the increase of $\varepsilon_{sp}$ leads to different behaviour in
the rugotaxis efficiency. Moreover, we have found that $A=0.5\sigma$ leads to the most
efficient rugotaxis, particularly for smaller droplets.

We take a closer look at the rugotaxis efficiency and inspect the different scenarios in 
our \textit{in silico} experiments, providing, also, further insight into the parameters
determining rugotaxis' performance (Fig.~\ref{fig:4}). We have identified a clear
distinction between substrates with small ($A=0.5\sigma$) and larger amplitudes $A>0.5\sigma$.
In particular, in the case of shallow wrinkles, $A=0.5\sigma$, we observe 
two different scenarios. In the first scenario,
the droplets attain their highest velocity $V$ for small $\lambda_0$, namely 
$\lambda_0=2\sigma$ and
$\lambda_E=\sigma$, albeit $V$ tends to decrease with growing adhesion $\varepsilon_{sp} 
\gtrsim  0.4$. In particular, the motion of small droplets ($N_p=100$) is 
significantly hampered as the
adhesion strength, $\varepsilon_{sp}$, grows.

\begin{figure*}[tb!]
\centering
  \includegraphics[width=0.9\textwidth]{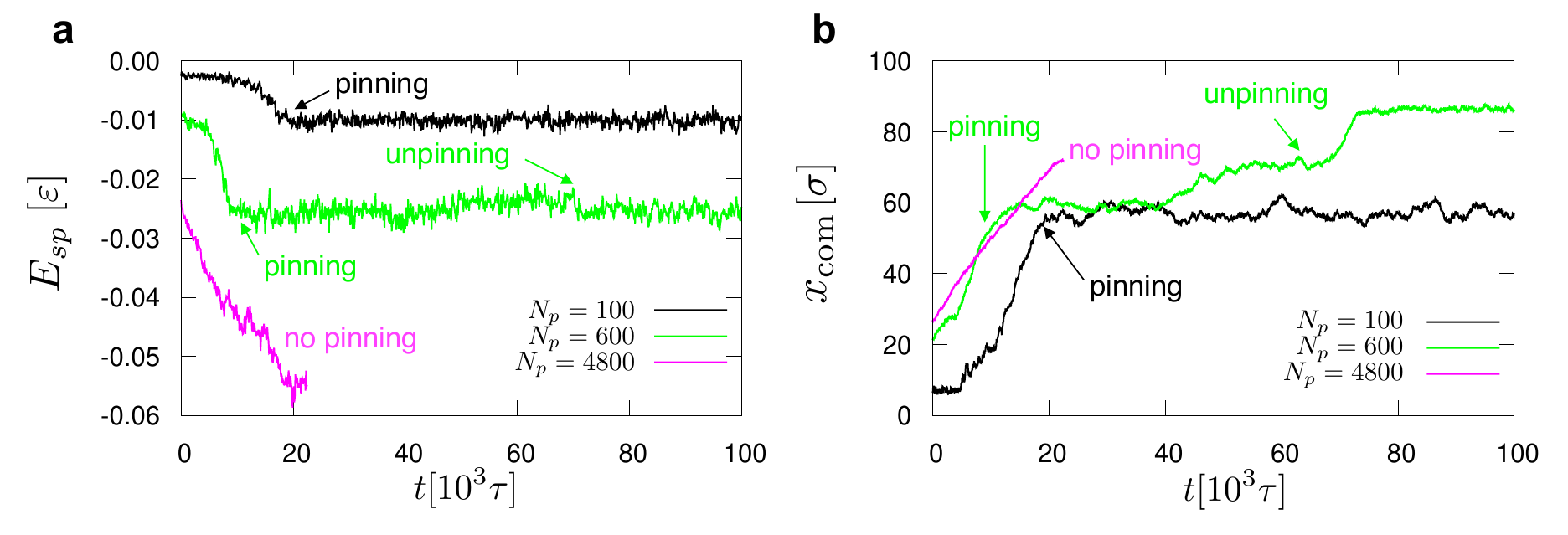}
\caption{\label{fig:5} Interfacial energy, $E_{sp}$, and position of the mass centre of
the droplet, $x_{com}$, as a function of time, as indicated. Hence, the distance 
covered by the droplet refers to its centre of
mass and droplets of different size will move a different 
distance in each \textit{in silico} expriment.
Arrows indicate the 
the time that pinning occurs for different cases of $N_p$ as indicated. 
We also indicate the unpinning of the droplet with an arrow in the case of $N_p=600$. For
$N_p=100$, the droplet remains pinned throughout the simulation. For
$N_p=4800$, there is no pinning. Here, $A=2\sigma$, $\lambda_0 = 4\sigma$, $\lambda_E=\sigma$,
$\varepsilon_{sp}=0.4 \varepsilon$. $N_p=100$, $600$ and $4800$, as indicated. }
\end{figure*}

In contrast (cf. Fig.~\ref{fig:4}a), larger gradients (\textit{e.g} 
$\lambda_0=6\sigma, \lambda_E=3\sigma$),
generally lower the mean velocity, $V$, which is found, however, to grow then 
steadily with increasing adhesion $\varepsilon_{sp}$. 
Moreover, the larger gradient ($\lambda_0=6\sigma, \lambda_E=2\sigma$) nearly doubles $V$ with
respect to ($\lambda_0=6\sigma, \lambda_E=3\sigma$) whereby for 
$\varepsilon_{sp}=0.6$ the speed of the largest drop attains that from the first scenario.
For deeper wrinkles, $A=2\sigma$ (Fig.~\ref{fig:4}b), the same trend is observed 
although pinning effects  occur more frequently so that for the case, $\lambda_0=2\sigma$,
$\lambda_E=\sigma$ (not shown here), no rugotaxis whatsoever takes place.
Apart from the smallest
droplets, $N_p=100$, an increase in the gradient $G_\lambda$ leads similarly to faster motion,
which becomes more efficient with growing drop size. However, further increase of 
the gradient or $\lambda_0$ will hinder the motion of the droplets, 
independently of the droplet size. Moreover values of 
$\varepsilon_{sp}=0.7\varepsilon$ (strong adhesion of 
the droplet to the substrate) will always prevent the motion of the droplet
or lead to complete droplet spreading (\textit{e.g.} for $N_p=4800$).
Still, we have observed that larger droplets overcome more easily the pinning barriers, 
which is in agreement with recent simulation results \cite{Theodorakis2021}.

As in the case of durotaxial motion \cite{Theodorakis2017}, we have identified the change in
the interfacial energy between the droplet and the substrate as the main driving force of
rugotaxis, also determining its efficiency (Fig.~\ref{fig:5}).
In our case, the interfacial energy is simply the total pair interaction
between the droplet and the substrate beads.
Crucially, for
a certain droplet size this depends on the particular parameters, $A$, 
$\varepsilon_{sp}$, $\lambda_0$, 
and $\lambda_E$. We have also identified cases when the droplet exhibits pinning during
rugotaxis but is also able to overcome the associated energy barrier caused by the pinning
and eventually complete the rugotaxial motion. While in such cases, rugotaxis is successful
and eventually the droplet reaches the very right end of the substrate (Fig.~\ref{fig:5}), 
in many other cases considered in our study as unsuccessful rugotaxis 
cases, the droplet 
failed to overcome the associated energy barrier.
Typically, such pinning events occur  at the transition from $W$- to $CB$-state 
as the drop partially climbs 
out of the substrate troughs whereby the interfacial energy $E_{sp}$ rises 
before the {\it whole} droplet 
attains a new $CB$-state where droplet beads cannot enter into the 
narrowed wrinkles\cite{Theodorakis2021}.

Figure~\ref{fig:5} contains plots of the position of the centre of mass of the droplet,
$x_{com}$, and the interfacial energy, $E_{sp}$, as a function of the elapsed time from the
beginning until the end of the rugotaxial motion. For $N_p=600$, 
we find that the droplet tranlocates
efficiently until pinning occurs.
An important point here is that the energy, $E_{sp}$, continues to decrease 
just after the pinning, which
allows the droplet to move to the right, where the underlying density (and wettability)
of the substrate increases 
with the wavelength gradient.

In summary, we have investigated solid, wrinkled substrates with and without wavelength
gradient in one direction, in an attempt to explore the rugotaxial motion of droplets. To
achieve our goal we have mainly employed MD simulation based on coarse-grained force-fields,
as we have done previously in the case of durotaxis.\cite{Theodorakis2017} We have also
provided relevant theoretical background and, also, performed numerical calculations based
on a hybrid MD--DFT approach. \cite{LoVerso2011,Milchev2018,Xu2007} 

We have constructed the diagram of W and CB states for a range of parameters. For small
wrinkle amplitudes, the W state prevails for a range of droplet--substrate adhesion
strength and wrinkle wavelengths. As the amplitude increases, the CB state sets in initially
at small adhesion strengths and wrinkle wavelengths. As the amplitude further increases,
the CB state appears for a larger range of wavelengths and adhesion strengths. Theory
\cite{Carbone2005} predicts the boundary between W and CB states, which is in agreement with
the simulation results. Based on a hybrid MD--DFT approach, we have identified the
interfacial energy-term to be mainly responsible for the different behaviour (W or CB).

In the case of the rugotaxis MD \textit{in silico} experiments, we find a range of 
optimal choices of the parameters that govern rugotaxis.
In the case 
of shallow wrinkles with $A=0.5\sigma$, we observe the most efficient rugotaxial 
motion for $\lambda_0=2\sigma$, which
corresponds to a small wavelength gradient.
For larger amplitudes, increased wavelength gradients lead to
more efficient rugotaxis up to some moderate adhesion strength.
As a result,
our results indicate that
an appropriate substrate design will determine the efficiency of the rugotaxial motion.

Based on our analysis, we have identified that changes of the interfacial energy between
the substrate and the droplet constitute the driving force of rugotaxial motion, as has 
been also found previously in the  case of durotaxis \cite{Theodorakis2017}. Hence, the two
phenomena share characteristics, but, in the case of rugotaxis, this crucially depends on 
a larger number of parameters that can lead to different scenarios, with pinning playing a 
crucial role. We anticipate that our
study elucidates important aspects of the rugotaxis phenomenon and suggests design
principles for nano-fabricated wirnkle-patterned substrates that can increase the efficiency
of rugotaxial motion.




\acknowledgments
This research has been supported by the National Science Centre, Poland, under grant
No. 2019/35/B/ST3/03426. A. M. acknowledges support by COST (European Cooperation in
Science and Technology [See http://www.cost.eu and https://www.fni.bg] and its 
Bulgarian partner FNI/MON under KOST-11). 
This research was supported in part by PLGrid Infrastructure.

\bibliographystyle{unsrt}

\end{document}